\documentclass[aps,prl,twocolumn,superscriptaddress]{revtex4}

\usepackage{graphicx}

\begin{document}

{\bf Comment on ``Memory Effects in an Interacting Magnetic Nanoparticle
System''}

In a recent Letter, Sun et al. \cite{sun} study and discuss memory effects in
an interacting nanoparticle system with specific temperature and field 
protocols. The authors claim that the observed memory effects originate 
from spin-glass dynamics and that the results are consistent with 
the hierarchical picture of the spin-glass phase. 
In this comment, we argue their claims premature by demonstrating 
that all their experimental curves can be reproduced qualitatively using only 
a simplified model of {\em isolated} nanoparticles~\cite{klik} with a
temperature dependent distribution of relaxation times. 

\begin{figure}[hb]
\includegraphics[width=0.85\columnwidth]{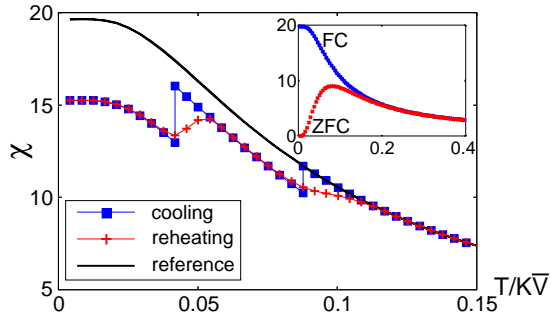}
\caption{FC susceptibility vs temperature using the
same protocol as in Fig.~2 of Sun et al.~\cite{sun}. The field is cut
during the temporary stops of the cooling at $T=0.088$ and at
$T=0.042$ for $10^{14}\tau_0$. The cooling (and reheating) rate is 
$2.4\times 10^{13}\tau_0$ per temperature unit. The inset shows ZFC and
FC susceptibility  vs temperature. \label{fig1}}
\end{figure}

The $i$-th magnetic moment in our model occupies one of two states
with energies $-K V_i\pm H M_{\rm s} V_i$, where $K$ is 
the anisotropy constant, $M_{\rm s}$ the saturation magnetization, 
$H$ the applied field and $V_i$ the volume of the $i$-th nanoparticle. 
The superparamagnetic relaxation time is $\tau_i=\tau_0\exp(KV_i/T)$.
The occupation probability of one of the states is $p_i(t)$, which is solved 
by the master equation approach  for any temperature and field protocol 
from a given initial condition \cite{klik}. 
The magnetization of the particle 
system is evaluated by averaging over the volume distribution 
$P(V)=\exp[-\ln(V)^2/(2\gamma^2)]/(\gamma V \sqrt{2 \pi})$ with 
$\gamma=0.6$. 

Figure~\ref{fig1} shows field-cooled (FC) magnetization vs
temperature measured on cooling---with temporary stops under zero
field---and the subsequent reheating. Since the field is cut 
at $T_{\rm s}$ for $t_{\rm s}$, $\{ p_i(t) \}$ of moments which are active on 
the present time scale relax to 1/2. Among them, 
moments of particles fulfilling 
$t_{\rm s}\approx \tau_0 \exp(KV_i/T_{\rm s})$ are frozen 
in certain values when the cooling is restarted. Those frozen states are
reactivated when the system is reheated to $T_{\rm s}$, causing a dip 
in $\chi$. The time evolutions of the thermo-remanent-magnetization (TRM)
 shown in Fig.~2 can similarly be understood; an energy barrier 
specifies quite sharply a
temperature, below (above) which the moment is blocked (superparamagnetic).

An appropriate protocol to confirm memory effects due to spin-glass
dynamics is a zero-field-cooled (ZFC) process with a stop during cooling
under zero-field~\cite{mathieu}. In  a spin glass the
correlation length of spin-glass order grows during the stop and a
memory dip shows up upon reheating, but not in a noninteracting
nanoparticle system. 
This protocol, however, has not been examined in \cite{sun}.
%

We have argued that a distribution of (free-)energy barriers is
a sufficient origin of the memory effects discussed in~\cite{sun}. 
In noninteracting nanoparticle systems the distribution of relaxation 
times originates only from that of the particle volumes,
and is thus extrinsic and static. In spin glasses, on the other
hand, it is the consequence of the cooperative nature of spins
with randomly frustrated interaction, and is intrinsic and
dependent on the age of the system.
To conclude, only through the memory effects studied by Sun et al.~\cite{sun}
we cannot draw any conclusion whether a nanoparticle system is 
a non-interacting superparamagnet or an interacting spin glass.

\begin{figure}[h]
\includegraphics[width=0.8\columnwidth]{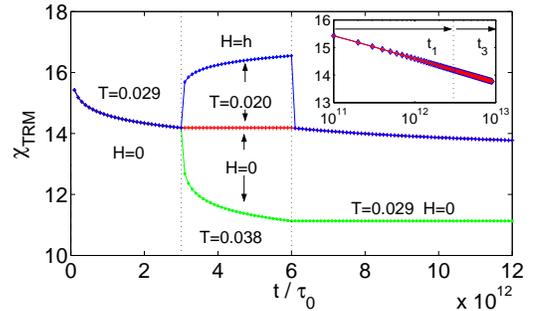}
\caption{$\chi_{\rm TRM}$ vs time using the same protocols as in
Figs. 3,4,5 of Sun et al.~\cite{sun}. The system is cooled to $T=0.029$
at the same rate as in Fig.~\ref{fig1} under a field which is cut just before
 recording $\chi_{\rm TRM}$. After a time of
$t_1=3\times10^{12}\tau_0$ the temperature is changed. 
The relaxation at the new temperature is recorded either in $H=0$ or $H=h$ 
in period of $t_2=3\times10^{12}\tau_0$. Then the temperature is shifted
back to $T=0.029$ and the field is set to zero. 
\label{fig2}}
\end{figure}

\acknowledgments
M.S. and P.E.J. acknowledge financial support from the Japan Society for the Promotion of Science.

\vspace*{.1cm}

\noindent
M. Sasaki,$^1$ P. E. J{\"o}nsson,$^1$ H. Takayama,$^1$ and P. Nordblad$^2$\\
{\small
$^1$ISSP, Univ. of Tokyo, Kashiwa, Chiba 277-8581, Japan \\
$^2$\AA-lab, Uppsala Univ.,
Box 534, SE-751 21 Uppsala, Sweden }

\end{document}